# Two-Dimensional Ferroelastic Topological Insulators in Single-Layer Janus Transition Metal Dichalcogenides MSSe (M=Mo, W)

Yandong Ma[†], Liangzhi Kou[⊥], Baibiao Huang[†], Ying Dai[†,*], and Thomas Heine[‡,§,*]

[†] School of Physics, Shandong University, Shandanan Str. 27, 250100 Jinan, People's Republic of China

[⊥] School of Chemistry, Physics and Mechanical Engineering Faculty, Queensland University of Technology, Garden Point Campus, QLD 4001, Brisbane, Australia

[‡] Theoretical Chemistry, School of Science, TU Dresden, Mommsenstr. 13, 01062 Dresden, Germany

[§] Wilhelm-Ostwald-Institut für Physikalische und Theoretische Chemie, Universität Leipzig, Linnéstr. 2, 04103 Leipzig, Germany

*Corresponding author: daiy60@sdu.edu.cn (Y.D.); thomas.heine@tu-dresden.de (T.H.)

Two-dimensional topological insulators and two-dimensional materials with ferroelastic characteristics are intriguing materials and many examples have been reported both experimentally and theoretically. Here, we present the combination of both features – a two-dimensional ferroelastic topological insulator that simultaneously possesses ferroelastic and quantum spin Hall characteristics. Using first-principles calculations, we demonstrate Janus single-layer MSSe (M=Mo, W) stable two-dimensional crystals that show the long-sought ferroelastic topological insulator properties. The material features low switching barriers and strong ferroelastic signals, beneficial for applications in nonvolatile memory devices. Moreover, their topological phases harbor sizeable nontrivial band gaps, which supports the quantum spin Hall effect. The unique coexistence of excellent ferroelastic and quantum spin Hall phases in single-layer MSSe provides extraordinary platforms for realizing multi-purpose and controllable devices.

TOC Figure

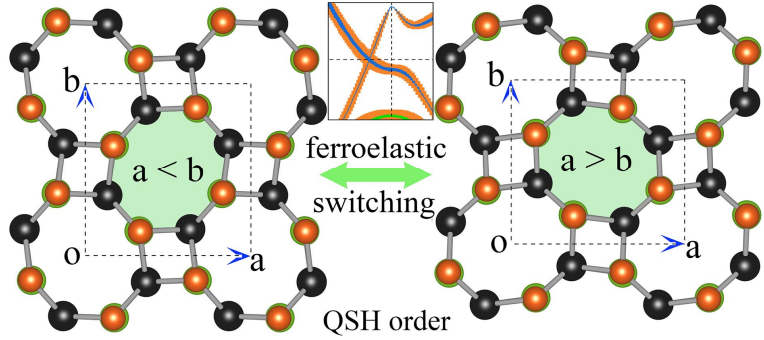

QSH order

## I. Introduction

Two-dimensional (2D) crystals can harbor a variety of fundamentally new physical properties that offer distinct features that may become essential for next-generation nanoscale devices [1,2]. Among these extraordinary properties, special emphasis has been given to 2D ferroelasticity [3,4] and quantum spin Hall (QSH) phase [5,6]. The defining signature of a 2D ferroelastic material is the existence of two or more equally stable geometries which can be interchanged by switching without diffusion of atoms by the application of external stress [6-8]. 2D ferroelasticity holds potential for applications in nonvolatile memory devices and acquires a growing importance. But so far only a few 2D ferroelastic materials have been reported, such as phosphorene, phosphorene analogues, t-YN, and Borophane [9-12]. QSH insulators are a novel quantum state of matter characterized by the gapless edge states inside the bulk gap caused by spin-orbit coupling [13,14]. As protected by time-reversal symmetry, charge carriers in such edge states are robust against backscattering, offering a fascinating way to energy-efficient electronic devices and spintronics [15]. The currently observed QSH effect is limited at very low temperature [16,17], and hence there is a great interest in searching for room temperature QSH insulators [18-22]. As spin-orbit coupling determines the size of the nontrivial band gap, the materials involved with heavy elements receive particular attention.

Compared with 2D materials possessing ferroelastic or QSH characteristics solely, undoubtedly, 2D ferroelastic topological insulators (FELTIs) that entail both features simultaneously are of particular interest as they will open up unprecedented opportunities for intriguing physics, whose exploitation may promise multi-purpose and miniaturized device applications [23-25]. Also, such a unique combination between ferroelastic and QSH orders holds appealing potential for controlling the anisotropy of the topological edge state via elastic strain engineering. Though 2D FELTIs are highly desirable, unfortunately up to date, there has been extremely limited work on it, largely owing to the fact that both 2D ferroelasticity and QSH insulator are rare themselves [9-12,16,17]. To our knowledge, the only existing potential candidate for realizing such 2D FELTIs is 1T′–WTe$_2$ as independent works demonstrated its ferroelastic [26] and nontrivial topological [27,28] properties separately. Switching from one ferroelastic state to the other in the 1T′ structure is associated by a reorientation of complete lines of Te atoms, subject to an interconversion barrier that we estimate to be 400 meV per unit cell. This, and the fact that homogeneous strain fields would not trigger the ferroelastic exchange, implies that this material is unlikely to perform as 2D FELTI. Thus the search for alternatives is vitally important for both fundamental scientific interest and practical applications.

In this work, using first-principles calculations, we computationally design a novel class of stable 2D materials, Janus single-layer (SL) MSSe (M=Mo, W). Using chemical vapor deposition (CVD), the 2H phase of SL MSSe has been prepared experimentally [29,30]. The corresponding distorted Haeckelite (S′) polytypes, whose stability we confirm on grounds of phonon dispersion calculations and Born criteria, inhibit the properties that characterize the long-sought 2D FELTIs. Switching inbetween their ferroelastic orders requires to overcome a barrier of 2.4 meV (MoSSe) and 10.2 meV (WSSe) per unit cell. The topological states are embedded in a bulk band gap of 29 meV (MoSSe) and 63 meV (WSSe) and we demonstrate their non-trivial topological character by calculating the $Z_2$ characteristics.

**II. Results and Discussion**

The crystal structure of SL MSSe (in the remainder of this work, we will only discuss single layers) adopts a distorted Haeckelite configuration [31-33] and is shown in **Figure 1(a)**. Similar as in the T′ polytype, a distorted T polytype which is known for MTe$_2$ (M = Mo, W) [27], we label this structure using S′. The structure shows PBA2 symmetry (space group no. 32) with a tetragonal lattice. The optimized lattice constants are *a*=6.34 (6.44) and *b*=6.64 (6.50) Å for 1S′–WSSe (MoSSe). In each unit cell [marked by the dashed line in **Figure 1(a)**], it consists of four M, four S and four Se atoms, with the M layer sandwiched between the S and Se layers. From the top view four- and eight-membered MSSe rings are observed. The basal planes are structurally similar to most TMDCs: they are chemically saturated and therefore rather non-reactive. To reflect the bonding character in MSSe, in **Figure 1(b)** and **S1(a)** we plot the electron localization function (ELF) map in the vertical plane containing two M, two S and two Se atoms. The electron localizations are mainly distributed at the M and chalcogen sites as well as around the centers between them, producing a covalent character of the M-S/Se bonding. Aside from the centers between M and chalcogen atoms, some electron localization also appears between two metal atoms, signifying a chemical bond is also formed between the two closest metal atoms. The crystal orbital Hamilton populations (COHPs) for the M-M bond, as well as the corresponding local densities of states at the metal centers (LDOS) [**Figure 1(d)** and **Figure S1(b)**] are very similar for both 1S′-MSSe, with populated antibonding states across the Fermi level, which suggests that the M-M bonding is rather weak. Overall integration of the COHP curves in the occupied levels, which are responsible for bond formation, gives values of 0.55 and 0.50 for 1S′–WSSe and 1S′–MoSSe, respectively, showing that the bonds in both compounds are comparable and neither of the bonding is strong.

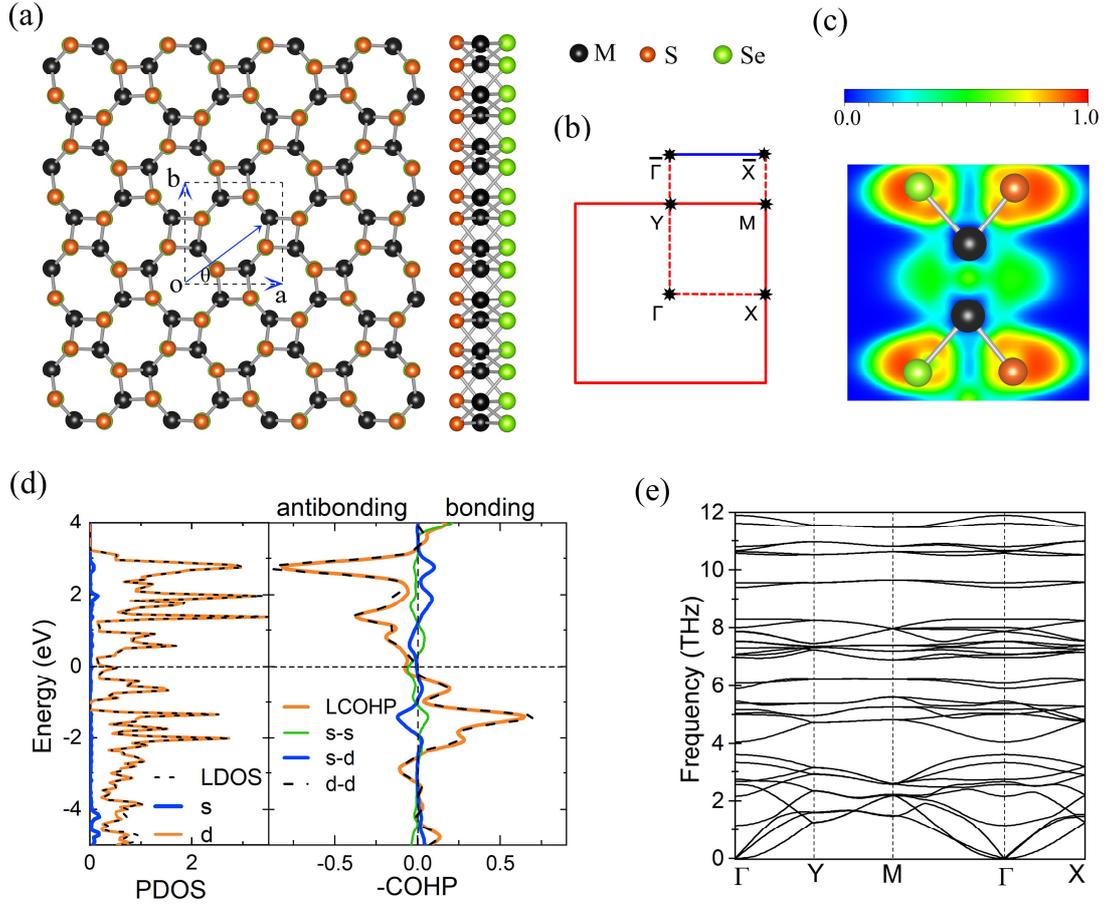

**Figure 1**. (a) Crystal structures of SL MSSe from top (left panel) and side (right panel) views, with the black square marking the primitive cell. (b) 2D and projected one-dimensional Brillouin zones with high-symmetry points. (c) Electron localization function (ELF) for 1S′–WSSe. (d) Projected density of states (PDOS) and crystal orbital Hamilton population (COHP) analysis for the W-d orbital projections of the nearest-neighbor W–W interactions. LDOS and LCOHP stand for the total DOS and the total COHP at the respective metal atoms. (e) Phonon spectra of 1S′–WSSe.

The dynamic stability of 1S′–MSSe is verified by calculating their phonon dispersion relations. As shown in **Figure 1(e)** and **S3(a)**, both structures are free from imaginary frequencies in the whole Brillouin zone, demonstrating that they are dynamically stable. The thermal stability of 1S′–MSSe is further validated by performing *ab initio* MD simulations using a 2 × 2 supercell (300K, 10 ps, for details see SI and **Figure S2, S3**), where no geometric rearrangement has been observed. For a mechanically stable 2D material, the elastic constants should obey the Born criteria: $C_{11}C_{22} - C_{12}^2 > 0$ and $C_{66} > 0$ [34,35], which are satisfied for proposed 1S′–WSSe (1S′–MoSSe) structures, as their calculated elastic constants are $C_{11}$ = 44.3 (49.7) N/m, $C_{22}$ = 69.9 (52.4) N/m, $C_{12}$=43.8 (43.7) N/m and $C_{66}$ = 37.9 (29.1) N/m. On grounds of phonon dispersion, molecular dynamics simulations and Born criteria we conclude that the 1S′–MSSe are structurally robust and potentially realizable in experiment.

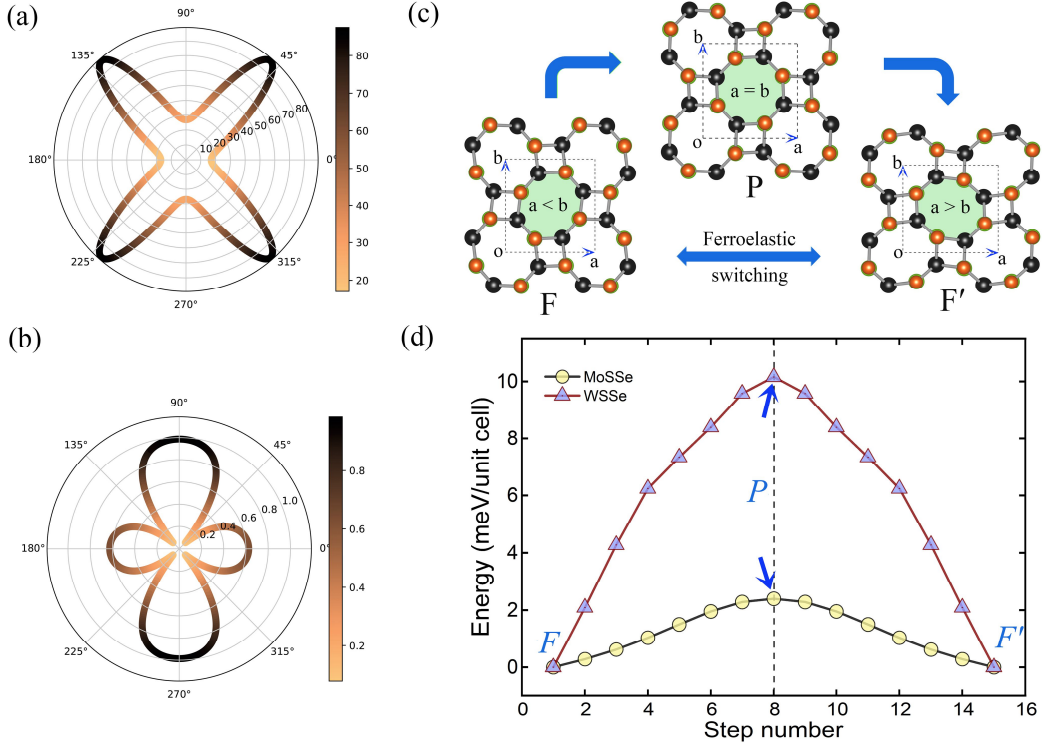

**Figure 2**. (a) Young's modulus and (b) Poisson's ration of 1S′–WSSe as a function of the angle θ. θ = 0º corresponds to the *a* axis. (c) Ferroelastic switching between two ferroelectric states F (left panel) and F′ (right panel) for 1S′–WSSe, with the transition state connecting them. (d) Energy pathway of the ferroelastic switching as a function of step number in the climbing image nudged elastic band (cNEB).

We now investigate those mechanical properties that are relevant to reveal ferroelasticity, i.e. the Young's modulus Y(θ) and Possion's ratio υ(θ), of 1S′–MSSe. Angle θ (Figure 1(a)), controls the interconversion between two ferroelstic orders. On the basis of the elastic constants, the Young's modulus Y(θ) and Possion's ratio υ(θ) along the in-plane θ can be expressed as follows [36]:

$$Y(\theta) = \frac{C_{11}C_{22}-C_{12}^2}{C_{22}\cos^4\theta+A\cos^2\theta\sin^2\theta+C_{11}\sin^4\theta},$$

$$\nu(\theta) = \frac{C_{12}\cos^4\theta-B\cos^2\theta\sin^2\theta+C_{12}\sin^4\theta}{C_{22}\cos^4\theta+A\cos^2\theta\sin^2\theta+C_{11}\sin^4\theta}.$$

Here, A = $(C_{11}C_{22} - C_{12}^2)/C_{66} - 2C_{12}$ and B = $C_{11} + C_{22} - (C_{11}C_{22} - C_{12}^2)/C_{66}$. The angular dependent results are shown in **Figure 2** and **S4**. The calculated Young's modulus for 1S′–WSSe (MoSSe) varies from 16.8 (13.2) N/m to 89.9 (72.5) N/m, indicating the mechanical anisotropy for both structures. These values are comparable to those of silicene (62 N/m) [37] and phoshorene (24-103 N/m) [38], but smaller than those of graphene (~340 N/m) [39] and indicative for the feasibility of apply stress for imposing ferroelastic transitions. Possion's ratio is used to describe negative ratio of the transverse strain to the corresponding axial strain, and most materials harbor Possion's ratios between 0 and 0.5 [40]. Intriguingly, the Possion's ratios for 1S′– WSSe (MoSSe) are computed to be 0.6 (0.8) and 1.0 (0.9) along *a* and *b* axis, respectively. Such significantly large Possion's ratios suggest sensitive structural response to external stress, which also is beneficial for ferroelasticity.

A necessary condition for a ferroelastic material is that the barrier of interconversion between the degenerate ground states is in the right order – not too small to prevent interconversion due to temperature effects, and not too high to allow for switching by application of stress. The two ferroelastic ground states F and F′ are shown in **Figure 2(c)**. If the initial state is chosen to be F, the lattice constant *a* is shorter than *b*. Upon applying an external stress along *a* axis, the shorter lattice switches to *b* axis, that is, 1S′–MSSe transforms into the ground state F′, which has lattice constants are |a′| = |b| and |b′| = |a|. Note that the described deformation without bond breaking imposes a lattice transformation that is equivalent to various symmetry operations, namely a 90° rotation, and two reflexions. The paraelastic state (P) would be found at the transition between the two ferroelastic states in 1S′–WSSe (MoSSe), where the lattice constants are *a* = *b*=6.49 (6.47) Å. Such paraelastic state would experience spontaneous lattice relaxations along both *a* and *b* axis, giving rise to the ferroelastic ground states F and F′.

To get a better physical picture for the ferroelastic switching in 1S′–MSSe, we calculate the transformation processes from initial state F to final state F′ employing the climbing image nudged elastic band (cNEB) method [41]. The transition state connecting F and F′ is, as expected, the high-symmetry paraelastic state S-MSSe, and no other metastable states are found in the interconversion [**Figure 2(d)**]. The calculated energy barrier is 10.2 (2.4) meV per unit cell for 1S′–WSSe (MoSSe), which is comparable with that of SL SnSe (9.2 meV per unit cell) [9]. Such low energy barriers produce the desirable possibility of fast ferroelastic switching in 1S′–MSSe upon imposing external stress, despite the fact that such switching generally subjects to the domain wall motion [25]. Another key factor that affects the ferroelastic performance is the reversible ferroelastic strain [(|b| / |a| - 1) ×100%], as it controls the signal intensity. The calculated reversible ferroelastic strain for 1S′–WSSe (MoSSe) is 4.7% (0.9%). In view of the fact that typical ferroelastic materials show reversible strain ranging from 0.5% to 3% [7,42], 1S′–MSSe, specially 1S′–WSSe, would exhibit a distinct signal of switching. Therefore, 1S′–MSSe are intrinsic ferroelactic materials holding potential applications in memory devices.

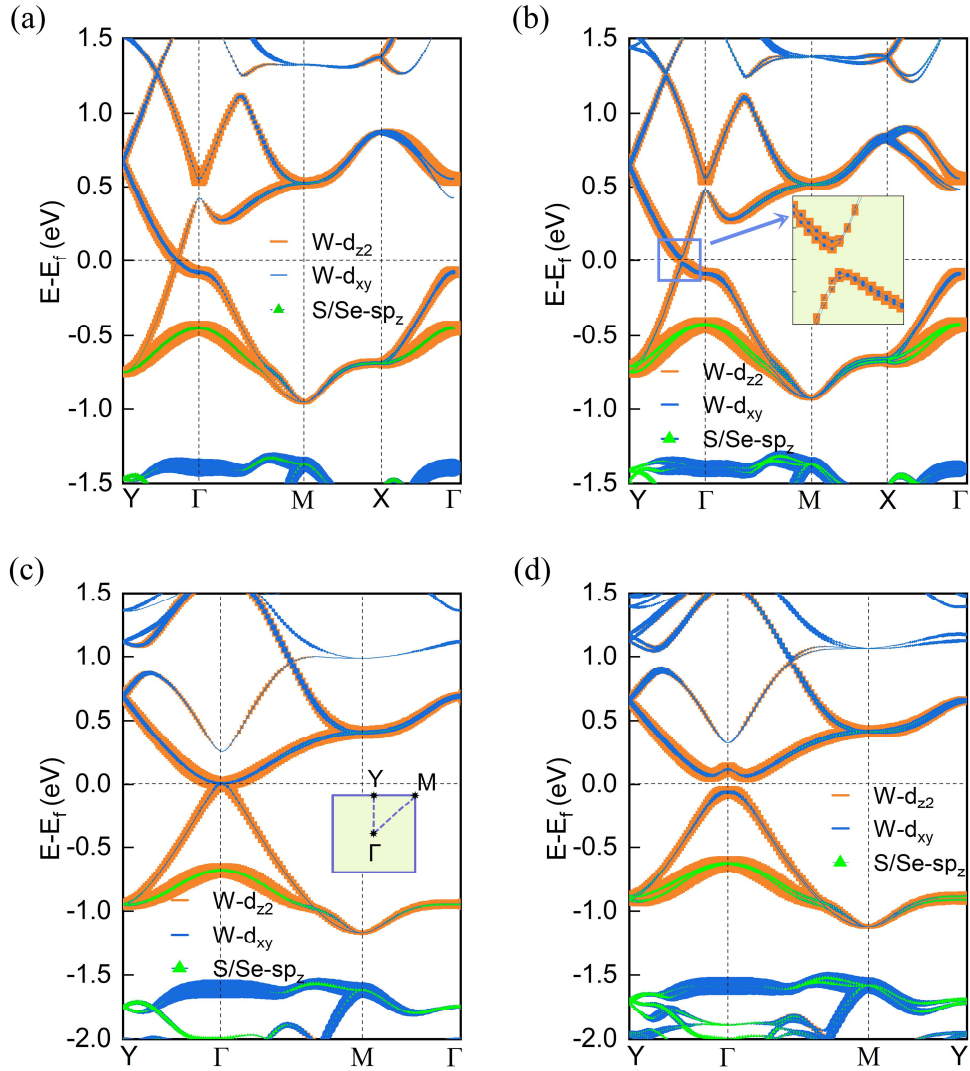

**Figure 3**. Fat band structures of SL WSSe (a) without and (b) with SOC. Fat band structures of SL WSSe in the paraelastic state (c) without and (d) with SOC. The Fermi level is set to 0 eV.

We next focus on the electronic properties of 1S′–MSSe: the fat bands are shown in **Figure 3** and **S3**. We can see that for both structures the bands in the vicinity of the Fermi level are mainly composed of M-$d_{z2}$ orbitals, with other orbitals staying away from the Fermi level. When switching off spin-orbit coupling (SOC), as shown in **Figure 3(a)** and **S5(a)**, the valence and conduction bands degenerate at the Fermi level with the crossing point lying close to the Γ point. Therefore, 1S′–MSSe are gapless semiconductors, akin to graphene. Upon introducing SOC, the degenerate M-$d_{z2}$ orbitals split into two states, with the Fermi level separating the valence and conduction bands; see **Figure 3(b)** and **S5(b)**. This leads to the insulating states in 1S′–WSSe and MoSSe with a global band gap of 38 and 17 meV, respectively. In 1S′–MoSSe, the weaker SOC strength of Mo as compared to that of W leads to the smaller gap opening. As PBE usually underestimates the band gap, we also adopt HSE06 to get more precise characterizations of the band structures (**Figure S6**).

The SOC-induced band gaps in 1S′–WSSe and 1S′–MoSSe using HSE06 are 63 and 29 meV, respectively. Apart from the difference in the gap values, the band dispersions for both systems based on HSE06 display similar characteristics as their corresponding PBE results.

For comparison, we also examine the electronic properties of 1S′–MSSe in the paraelastic phase (p-MSSe). Since the situation in p-MoSSe is very similar to that in p-WSSe, here we only discuss p-WSSe as an example. The calculated band structures of p-WSSe are plotted in **Figure 3(c)** and **(d)**, in which we can see that p-WSSe is also a gapless semiconductor in the case without SOC and transforms into insulating state by turning on SOC. In addition, the bands of p-WSSe near the Fermi level are also mainly constituted by the W-$d_{z^2}$ orbitals. Such scenario of band compositions and evolutions in p-WSSe are comparable to that in 1S′–WSSe. However, when excluding SOC, the two bands involved in the band crossing exhibit a parabolic behavior in p-WSSe, while it is linear for 1S′–WSSe. Moreover, the crossing points lies exactly at, instead of close to, the Γ point. These discrepancies are directly connected to the lattice distortion in p-WSSe. Interestingly, by transforming ferroelastic into paraelastic state, the band crossing of 1S′–WSSe would shift to the Γ point.

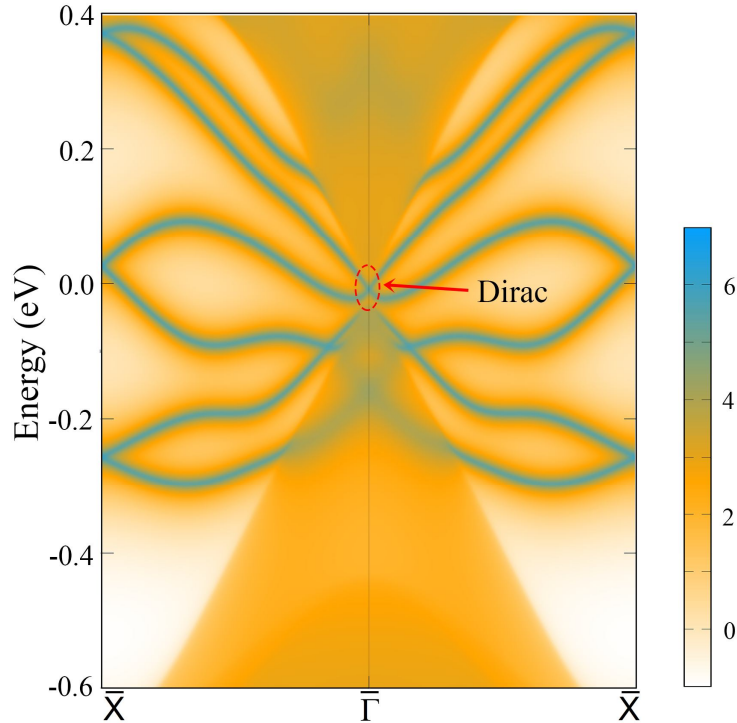

**Figure 4**. Local density of states of the semi-infinite SL WSSe. The Fermi level is set to 0 eV.

Such a SOC-induced phase transition from gapless semiconductor to insulator usually suggests the existence of nontrivial topological phase. To identify the nontrivial topological nature of 1S′–MSSe, we calculate the $Z_2$ invariant as a direct evidence. Using the maximally-localized

Wannier functions, $Z_2$ is obtained to be 1 for both structures, confirming 1S′–MSSe are QSH insulators. It is widely known that the hallmark of a QSH insulator is the appearance of odd-number of Dirac-like edge states connecting the valence and conduction bands. We then check the nontrivial topological edge states in 1S′–MSSe. With the maximally-localized Wannier functions, the edge Green's function of the semi-infinite 1S′–MSSe is constructed to calculate the edge local density of states. The local density of states of the semi-infinite 1S′–MSSe are shown in **Figure 4** and **S7**. As expected, there are a pair of edge states that connects the valence and conduction bands and produces one Dirac-like point at the boundary of the Brillouin zone, confirming the $Z_2$ analysis. Therefore, remarkably, 1S′–MSSe are 2D FELTIs that can support the ferroelastic and QSH orders simultaneously. We also note that the edge states are protected by the time-reversal symmetry and thus would always exist, but their details would depend on the edges. Accordingly, 1S′–MSSe hold appealing potential for controlling the anisotropy of the topological edge state via elastic stress engineering.

### III. Conclusion

In summary, using first-principles calculations, we demonstrate a novel class of 2D FELTIs in SL MSSe, which are thermally, dynamically and mechanically stable. In particular, both structures display excellent ferroelasticity with low switching barriers and strong ferroelastic signals, making them appreciate for nonvolatile memory device applications. They also exhibit intriguing topological nature with sizeable nontrivial band gaps, which enables their QSH effect. Most remarkably, in addition to fundamental interests, the unique coexistence of ferroelasticity and QSH insulator in SL MSSe offers the opportunities for achieving multipurpose and controllable devices. We wish to emphasize that SL MSSe are the first class of 2D materials ever reported displaying the ferroelasticity and QSH insulator simultaneously, despite the fact that independent works separately demonstrated the ferroelastic and topological order in 1T′–WTe$_2$. Our work will inspire further researches to explore similar effects and novel device applications.

### Methods

First-principles calculations are performed using VASP, with a plane wave basis (cutoff energy was set to 500 eV) and the projector augmented wave (PAW) [43-45] approach. For geometry optimizations, exchange and correlation interaction between electrons is treated by generalized gradient approximation as specified by Perdew, Burke and Ernzerhof (PBE) [46]. Heyd-Scuseria-Ernzerhof (HSE06) hybrid functional [47] is employed for the band structure calculations in order to overcome the problem of band-gap underestimation in PBE functional. For more details on the computational details, see the Supporting Information. $Z_2$ invariants are calculated by the Wannier90 code [48,49], in which a tight-binding Hamiltonian with the maximally-localized Wannier functions are fitted to the first-principles band structures.

**Supporting Information**

The results of computational method, ELF, COHP, phonon spectra, results of the MD simulation, mechanical properties, band structures and LDOS of 1S′–MoSSe, as well as band structures of 1S′–MSSe at the HSE06 level are given in the Supporting Information. This material is available free of charge via the Internet.

**Acknowledgement**

Financial support by the Deutsche Forschungsgemeinschaft (DFG, HE 3543/27-1), National Basic Research Program of China (Grant No. 2013CB632401), National Natural Science Foundation of China (Grant No. 11374190). We also thank the Taishan Scholar Program of Shandong Province and Qilu Young Scholar Program of Shandong University. Computer time at ZIH Dresden is gratefully acknowledged.